\documentclass[12pt]{article}

\usepackage{graphicx}
\usepackage{amsmath}
\begin{document}

\begin{center}
{\Large{Parametric oscillator in a Kerr medium: evolution of coherent states}}\\
R. Rom\'an-Ancheyta$^{(1)}$, M. Berrondo$^{(2)}$ and J. R\'ecamier$^{(1)}$\\
$^{(1)}$Instituto de Ciencias F\'isicas,
Universidad Nacional Aut\'onoma de M\'exico\\
C.P. 62210 Cuernavaca, Morelos, M\'exico

$^{(2)}$Dept. Physics and Astronomy, Brigham Young University\\
Provo, UT 84602, USA

ricardo@fis.unam.mx, \ berrondo@byu.edu, \ pepe@fis.unam.mx

\end{center}

 \begin{abstract}
 We study the temporal evolution of a coherent state under the action of a parametric oscillator and a nonlinear Kerr-like medium.  We make use of the interaction picture representation and use an exact time evolution operator for the time independent part of the Hamiltonian. We approximate the interaction picture Hamiltonian in such a way as to make it a member of a Lie algebra. The corresponding time evolution operator behaves like a squeezing operator due to the temporal dependence of the oscillator's frequency. We analyze the probability amplitude and the auto correlation function for different Hamiltonian parameters and we find a very good agreement between our approximate results and converged numerical calculations. 
  \end{abstract} 
   
   \noindent Keywords\\
   Parametric; Kerr; Temporal evolution

   \section{Introduction}
   Coherent states were introduced by Schr\" odinger in 1926 since the early stages of quantum mechanics \cite{schroedinger}. These quantum states are characterized by the fact that the trajectory of the center of the coherent wave packet evolves in time in the same way as a classical harmonic oscillator and its dispersion takes the minimum value allowed by  Heisenberg's principle. These facts make them especially useful as a link between the classical and the quantum theories.
      
Much later, in 1963, Glauber introduced the {\em field} coherent states, that is, coherent states for the electromagnetic field.  These states play an important role in quantum optics\cite{glauber63,glauber63a}. The development of lasers made it possible to prepare light fields which are very close to the coherent states. Glauber showed that the field coherent states can be obtained from any one of the three mathematical definitions: (i) as the right hand eigenstates of the boson annihilation operator $\hat a |\alpha\rangle = \alpha|\alpha\rangle$ with $\alpha$ a complex number, (ii) as those states obtained by application of the displacement operator upon the vacuum state of the harmonic oscillator $D(\alpha)|0\rangle = |\alpha\rangle$ with $D(\alpha)= \exp(\alpha \hat a^{\dagger}-\alpha^{*} \hat a)$, and (iii) as the quantum states with a minimum uncertainty product $(\Delta p)(\Delta q) =\hbar/2$ with $\Delta q = \Delta p$. The coherent states obtained from any one of these definitions are identical when one makes use of the harmonic oscillator algebra. Subject to  a linear interaction a coherent state evolves into a new coherent state, that is, they show temporal stability \cite{klauder,reca02}. Nieto and Simmons \cite{nieto1,nieto2,nieto3} generalized the notion of coherent states for potentials different from the harmonic oscillator with unequally spaced energy levels such as the Morse potential and the P\"oschl-Teller potential. The resulting states are localized, follow the classical motion and disperse as little as possible in time.  Gazeau and Klauder \cite{gazeau} proposed a generalization for systems with one degree of freedom possessing discrete as well as continuous spectra. These states present continuity of labeling, a resolution of the identity and temporal stability. Man'ko and collaborators \cite{manko1} introduced coherent states of an $f$-deformed algebra as eigenstates of a deformed annihilation operator $\hat A=\hat a f(\hat n)$ where $f(\hat n)$ is a function of the number operator that specifies the deformation. These states present nonclassical properties like squeezing and antibunching. The properties of their even and odd combinations have also been studied \cite{roy,recjau}.

In the presence of a nonlinear interaction, field coherent states evolve into non classical states. This can be achieved experimentally by passing a coherent state through a Kerr medium resulting in the appearance of distinguishable macroscopic superpositions of coherent states, the so called cat states \cite{yurke,haroche}.

The parametric harmonic oscillator, namely a harmonic oscillator with a time dependent frequency, has been studied from several points of view: using the method of adiabatic invariants \cite{lewis1,lewis2,lewis3,manko2,manko2a,manko2b}, super symmetric quantum mechanics \cite{ocasta}, algebraic methods \cite{micha,micha2}, and different approximation methods \cite{kiss}.    A particularly relevant realization of the parametric oscillator is  cavity quantum electrodynamics (CQED) where the frequency of a given field mode in the cavity can change in time due to the motion of the cavity walls or to changes in the dielectric function of the medium \cite{dodonov-klimov}. For instance, Wineland et. al. \cite{wineland} analyzed both theoretically and experimentally the loss of coherence caused by fluctuations in the trap parameters and in the amplitude and frequency of the laser beams, heating due to collisions with background gas, internal state decoherence due to radiative decay, and coupling to spectator levels.

 In this work we consider a nonlinear system corresponding to a single mode field propagating in a Kerr-like medium immersed in a cavity with a time dependent frequency. In Section \ref{sec-theory} we write the Hamiltonian and construct its time evolution operator. In section \ref{sec-stats} we follow the evolution of coherent states under the nonlinear Hamiltonian and analyze some of their statistical properties.   
   \section{Theory}
   \label{sec-theory}
Consider a parametric harmonic oscillator immersed in a Kerr-like medium. 
Its Hamiltonian is given by:
\begin{equation}
\hat{H}(t)= \frac{1}{2}[\hat{p}^2 + \Omega^2(t)\hat{q}^2] + \hat{H}_{Kerr}
\end{equation}
where $\Omega(t)$ is an explicit time dependent frequency and
$\hat{H}_{Kerr}$ has to do with the Kerr-like medium. 
We can define the usual annihilation, creation and number operators as:
\begin{equation}
\hat{a}=\frac{1}{\sqrt{2\Omega_0}}(\Omega_0\hat{q}+i \hat{p}),\ \ \
\hat{a}^\dagger=\frac{1}{\sqrt{2\Omega_0}}(\Omega_0\hat{q} -i \hat{p}),\ \ \
\hat{n}=\hat{a}^\dagger\hat{a}.
\end{equation}
where we have set $\hbar=1$ and we write the Kerr medium  \cite{milburn} as
$\hat{H}_{Kerr}=\chi\hat{n}^2$, with $\chi$ a constant proportional to a
third-order nonlinear susceptibility $\chi^{(3)}$ which is, in general, a small number \cite{boyd}.
To be specific, in what follows we will choose 
$\Omega(t)= \Omega_0[1+2\kappa \cos(2\Omega_0 t)]$ \cite{manko58}
with $\kappa$ also a small parameter.
The Hamiltonian can be written in terms of $\hat{a}^\dagger$,
$\hat{a}$ and $\hat{n}$ as:
\begin{eqnarray}\label{eq:hnl}
\hat{H}(t) & = \Omega_0(\hat n + 1/2) + \chi \hat n^2 + g(t) (\hat a^2 + \hat a^{\dagger 2} + 2\hat n +1)
\end{eqnarray}
and $g(t)=\Omega_0 \kappa \cos(2\Omega_0 t)(1+\kappa \cos(2\Omega_0 t))$.

The time evolution operator corresponding to the non linear time independent part of the Hamiltonian is given by:
\begin{equation}\label{eq:opevol0}
\hat{U}_0 = \exp\left(-i\Omega_0 t(\hat n+1/2) -i t \chi  \hat n^2\right)
\end{equation}
and we can write the time dependent Hamiltonian in the interaction picture as
\begin{equation}\label{eq:HI} 
\hat{H}_I(t)= g(t) \left(e^{-2 i\Omega(\hat n) t} \hat a^2 +
\hat a^{\dagger 2}e^{2i\Omega(\hat n) t} + 2\hat n +1\right)
\end{equation}
where we have used the fact  that $ f(\hat n) \hat a = \hat a f(\hat n-1)$
and $f(\hat n)\hat a^{\dagger} = \hat a^{\dagger}f(\hat n+1)$ and the
effective frequency $\Omega(\hat n) = \Omega_0+2\chi(1+\hat n)$ 
is a function of the number operator.

 Notice that the time evolution operator $\hat{U}_0$ is exact and includes the anharmonicity due to the Kerr medium explicitly. The interaction picture Hamiltonian given in Eq.~\ref{eq:HI} is also exact. The operators given in the interaction picture Hamiltonian do close under commutation, however they have an explicit  time dependence and the Wei-Norman theorem can not be applied. Nevertheless  the set $\{\hat a^{\dagger 2}, \hat n+1/2, \hat a^2\}$ also forms the basis 
of a Lie algebra (the $su(1,1)$ algebra) closed under commutation. In order to attain a more manageable Hamiltonian that can be written as a linear combination of time independent operators we approximate the exponentials by their average value \cite{MBJR}, that is, we make the replacement $\exp[\pm 2i \Omega(\hat n) t]$ by 
$\langle\alpha_0|\exp[\pm 2i \Omega(\hat n) t]|\alpha_0\rangle$
 obtaining the approximate interaction picture Hamiltonian:
\begin{equation}\label{eq:hint}
\tilde H_I(t)= g(t)\left( e^{-2i(\Omega_0+2\chi)t}\hat a^2\langle e^{-4 i\chi t(\hat n)}\rangle  +
\hat a^{\dagger 2} e^{2i(\Omega_0+2\chi)t}\langle e^{4 i\chi t(\hat n)}\rangle
+2\hat n +1\right)
\end{equation}
where the expectation value is taken with respect to an initial coherent state.
The resulting approximate Hamiltonian is similar to that of a degenerate parametric amplifier \cite{milburn}, where a non linear medium is pumped by a strong laser inducing the emission and absorption of photon pairs \cite{knight}.
 
With this simplification $\tilde H_I(t)$ is an element of the Lie algebra with time dependent coefficients and the corresponding time evolution operator may be written {\em exactly} in the product form \cite{wei-norman,wei2}
\begin{equation} \label{eq:UI}
\tilde H_I(t) = \sum_{n=1}^{4} f_n(t) \hat{X}_n , \ \ \ 
\hat{U}_I(t) = \prod_{n=1}^{4} e^{\alpha_n(t) \hat{X}_n}. 
\end{equation}
with initial conditions $\alpha_n(t_0)=0$, 
and we have chosen the ordering
$\hat{X}_1=\hat a^{\dagger 2}$, $\hat{X}_2=\hat n$, $\hat{X}_3=\hat a^2$ and $\hat{X}_4=1$.

The average takes the form:
\begin{equation}
e^{\pm2i(\Omega_0+2\chi)t}\langle \alpha_0|e^{\pm4i \chi t \hat n}|\alpha_0\rangle =e^{\pm 2i(\Omega_0+2\chi)t}\exp[|\alpha_0|^2(e^{\pm 4i \chi t}-1)].
\end{equation}
The complex, time dependent functions $\alpha_n(t)$ needed to construct $\hat{U}_I(t)$ are obtained from the following set of coupled, nonlinear, ordinary differential equations obtained after substitution of Eq.~\ref{eq:UI} in Schr\"odinger's equation
\begin{eqnarray}
\dot{\alpha_1} &=& -i(f_1 +2\alpha_1 f_2 +4 \alpha_1^2 f_3) \\
\dot{\alpha_2} &=& -i(f_2+4\alpha_1 f_3) \nonumber \\
\dot{\alpha_3} &=& -i f_3 e^{2\alpha_2} \nonumber \\
\dot{\alpha_4} &=& -i(f_4 +2\alpha_1 f_3)\nonumber
\end{eqnarray}
where the dot means the time derivative.
These equations can be solved either analytically or numerically, the equation for $\alpha_1(t)$ being a Riccati equation and the equations for the other $\alpha's(t)$ can be obtained by integration.

\section{Statistical properties}
\label{sec-stats}
\subsection{Probability distributions}
Once we have the explicit form for the time evolution operator, we can evaluate the temporal evolution of a coherent state $|\alpha\rangle$ by means of
\begin{equation}
|\alpha;t\rangle = \hat{U}_0(t) \hat{U}_I(t) |\alpha\rangle 
\end{equation}
which is given explicitly as:
\begin{equation}\label{eq:tdepcoh} 
|\alpha;t\rangle=N_{\alpha}
\sum_{l,m}^\infty\frac{(\alpha e^{\alpha_2-i\Omega_0 t})^l
(\alpha_1e^{-2i\Omega_0 t})^m}
{m!l![(l+2m)!]^{-1/2}}
e^{-i\chi t(l+2m)^2}|l+2m\rangle
\end{equation}
with
\begin{equation}
N_{\alpha}=
\exp\left(-i\Omega_0 t/2+\alpha_4+ \alpha^2\alpha_3-|\alpha|^2/2
\right).\nonumber
\end{equation}
The probability of finding the $k'$th excited state in the distribution at time $t$ is given by $P_{k}(\alpha;t)=|\langle k|\alpha;t\rangle|^2$. We obtain:
\begin{equation}
P_k(\alpha;t)=\left |N_{\alpha}\sqrt{k!}
\sum_{m=0}^{[k/2]}
\frac{\alpha_1^m (\alpha e^{\alpha_2})^{k-2m}}{m!(k-2m)!}
\right |^2
\end{equation}
where $[\gamma]$ means the integer part of $\gamma$.

\begin{figure}[h!]
\begin{center}
\includegraphics[scale=.8]{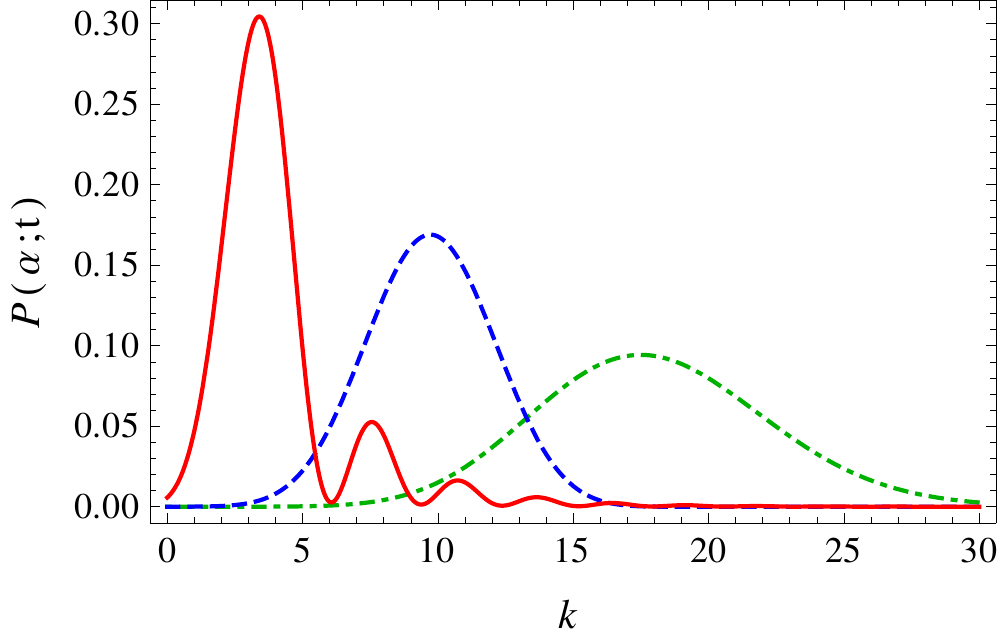}
\caption{Probability distribution for the evolved coherent state with Hamiltonian parameters $\chi=0$, $\kappa=0.05$, $\alpha=3+i3$, 
$\alpha_0=\sqrt{18}$ at times $t=0$ (green), $t=2\pi$ (blue) and $t=6\pi$ (red).}
\label{Dist_Prob_1}
\end{center}
\end{figure}

In Fig.~\ref{Dist_Prob_1} we show the probability distribution as a function of $k$ for three different values of time for the case when the Kerr term $\chi=0$ corresponding to a parametric harmonic oscillator. Here the  evolution due to the time independent part of the Hamiltonian is that of a harmonic oscillator and the interaction picture Hamiltonian is a linear combination of the operators $\{\hat a^2, \hat a^{\dagger 2}, \hat n\}$ so that it's time evolution operator is similar to a squeezing operator $S=\exp[\frac{1}{2}(\zeta^{*}\hat a^2 - \zeta \hat a^{\dagger 2})]$.\\
The Hamiltonian parameters used in this example are $\kappa=0.05$, $\alpha=3+i3$ and times $t=0$ (green), $t=2\pi$ (blue) and $t=6\pi$ (red). At the initial time the probability distribution is a Poissonian centered at $\langle n\rangle=18$ as corresponds to a usual coherent state. At $t=2\pi$ its width has decreased and it is now centered  at $k=10$ finally at $t=6\pi$ its maximum is located at $k=3$, its width is even smaller and it presents noticeable oscillations after an initial bell shape. These oscillations are evidence of the non classicality of the state and are due to the quantum interferences in phase space.

\subsection{Auto correlation function}
 The auto correlation function is defined as the overlap  \cite{Robinett}
\begin{equation}
F(t)=\langle \Psi(0)|\Psi(t)\rangle
\end{equation} 
 and it takes large values at times whenever the wave packet resembles the original one. When the overlap is complete we have a complete revival otherwise we may have fractional revivals when the overlap is a  fraction $(1/q)$ of the total probability. The phenomenon of wave packet revivals (complete or fractional) has been observed in many experimental situations in atomic and molecular systems \cite{yeazell, vrakking}.\\
Using the explicit forms of the time evolution operators $\hat{U}_0$ and $\hat{U}_I$, the time dependent coherent state can be expanded in terms of the number eigenkets $|n\rangle$ as given by Eq.~\ref{eq:tdepcoh} and the corresponding  auto correlation function is:
\begin{equation}\label{eq:autocorr}
F(t)=e^{- i\Omega_0 t/2+\alpha_4+z^2\alpha_3-|z|^2}
\sum_{k,l,=0}^\infty\frac{(|z|^2 e^{\alpha_2- i\Omega_0 t})^k}{k!}
\frac{(z^{*2}\alpha_1 e^{-i 2\Omega_0 t})^l}{l!}
 e^{-i\chi t(k+2l)^2}.
\end{equation}

 \begin{figure}[h!]
 \begin{center}
\includegraphics[scale=0.45]{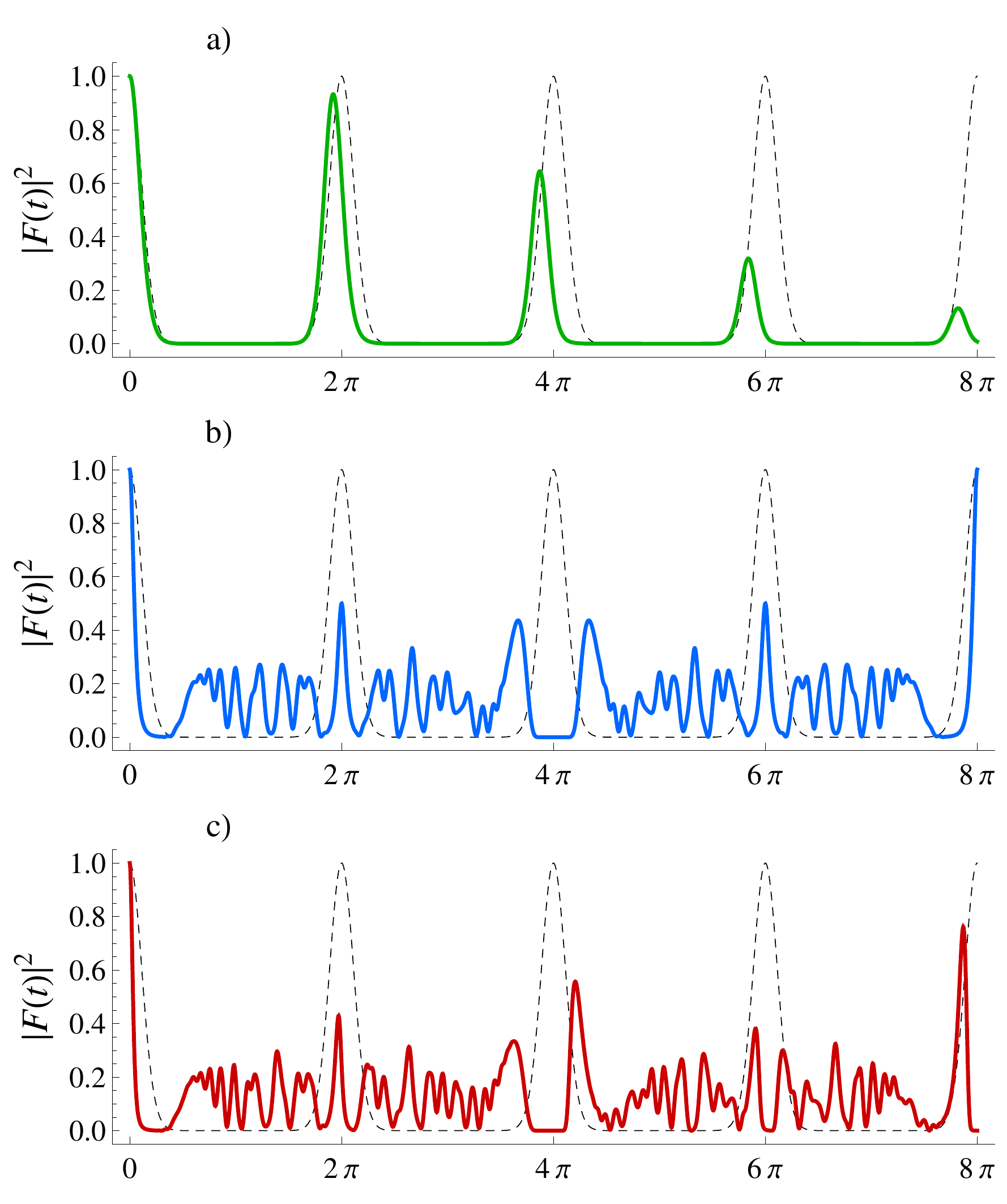} 
 \caption{
 Absolute value squared of the auto-correlation function $|F(t)|^2$ as a function of time
 for a state $|z;t\rangle$. Hamiltonian parameters,
$a)$: $\kappa=0.05$, $\chi=0$; $b)$: $\kappa=0$, $\chi=0.25$;
$c)$: $\kappa=0.25$, $\chi=0.25$. 
The black dotted line is for a field coherent state $|z e^{i\Omega_0 t}\rangle$.
In all cases we have $z=2$, $\alpha_0=2$ and $\Omega_0=1$.}
 \label{figautocorr}
 \end{center}
 \end{figure}

When the coefficient of the non linear term $\chi$ vanishes we deal with a parametric oscillator (Fig.\ref{figautocorr} top) and the auto correlation function for the coherent state $|z;t\rangle$ is a periodic decreasing function of time and its explicit form can be written as an  exponential
\begin{equation}
F(t)_{\chi=0}=\exp(-i\Omega_0 t/2+\alpha_4+z^2\alpha_3
-|z|^2+z^{*2}\alpha_1 e^{-i 2\Omega_0 t}
+|z|^2 e^{\alpha_2-i\Omega_0 t})	.
\end{equation}
As a reference we show in black the temporal evolution for a field coherent state $|z e^{i\Omega_0 t}\rangle$. 

When the coefficient corresponding to the temporal dependence of the frequency $\kappa\ll 1$ and that of the nonlinear term is non negligible (and we thus have a nonlinear oscillator) we show in Fig.~\ref{figautocorr} (intermediate) the auto correlation function for a coherent state $|z;t\rangle$. We can see that there are periodic fractional and complete revivals with the revival time $T_{rev}=4\pi\hbar/|E''(n_0)|=8\pi$. Notice also that the auto correlation function is symmetric with respect to $T_{rev}/2=4\pi$. In Fig.~\ref{figautocorr} (bottom) we show the case when neither $\kappa$ nor $\chi$ are negligible;  we have a parametric nonlinear oscillator. Here the auto correlation function shows only fractional revivals. The revival near  $8\pi$ is not complete and appears at a time slightly earlier than $T_{rev}$. Notice also that the periodicity with respect to $t=4\pi$ has been lost.

\begin{figure}[h!]
 \begin{center}
\includegraphics[scale=0.45]{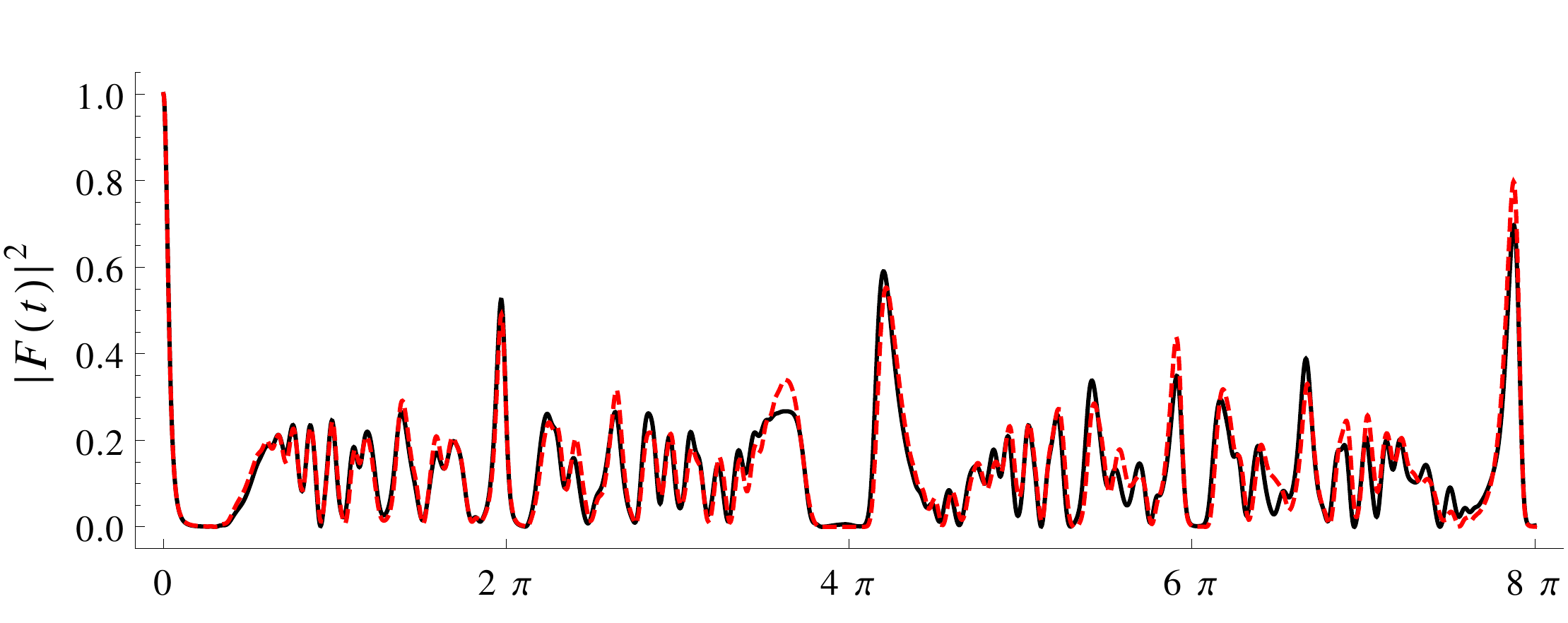} 
 \caption{
 Absolute value squared of the auto-correlation function $|F(t)|^2$ as a function of time
 for a state $|z;t\rangle$.
 Converged numerical results in black and our approximate results in red.
 Hamiltonian parameters: $\kappa=0.25$, $\chi=0.25$. In all cases we have $z=2$, $\alpha_0=2$ and $\Omega_0=1$.}
 \label{figautocorr2}
 \end{center}
 \end{figure}

In Fig.~\ref{figautocorr2} we plot the  absolute value squared of the auto-correlation function $|F(t)|^2$ as a function of time for a state $|z;t\rangle$ with Hamiltonian parameters $\kappa=0.25$, $\chi=0.25$. In red we present the result obtained using the time evolution operator $\hat{U}_I(t)$ obtained from the approximate interaction picture Hamiltoinian $\tilde H_I(t)$  and in black the result obtained when the evolution of the system is done numerically taking into account the full Hamiltonian given by Eq.~\ref{eq:HI}. Notice the almost perfect agreement between the converged numerical result and our approximate result. This is an indication of the quality of our approximation when dealing with the interaction picture Hamiltonian. We stress the fact that the nonlinearity $\chi \hat n^2$ has been taken into account {\em exactly} by means of the time evolution operator $\hat{U}_0$.

 \section{Conclusions}
 \label{sec-conclusions}
 In this work we have built an approximate time evolution operator for a system composed of a parametric oscillator in a nonlinear Kerr-like medium. The Hamiltonian is transformed into the interaction picture and as a result we obtained a time dependent Hamiltonian that contains the number operator in an exponential. In order to have a more managable Hamiltonian we approximate the exponential by its average value taken between  a time independent coherent state. With this simplification we can write the Hamiltonian in the interaction picture as an element of a finite Lie algebra whose time evolution operator can be expressed as a product of exponentials. To show the quality of our methodology we calculated probability distributions and the auto correlation function for a case where neither $\kappa$ nor $\chi$ are negligible. We found that the approximate method is consistent with the converged numerical results.
\vspace{.2cm}
  
\leftline{\bf Acknowledgements}
We thank Reyes Garc\'{\i}a for the maintenance of our computers and acknowledge partial support from CONACyT through project 166961 and DGAPA-UNAM project IN108413. One of us (MB) would like to thank the Instituto de Ciencias F\'{\i}sicas for its hospitality.


\end{document}